# dSDiVN: a distributed Software-Defined Networking architecture for Infrastructure-less Vehicular Networks

Ahmed Alioua[1], Sidi-Mohammed Senouci[2], Samira Moussaoui[1]

[1] Computer Science Department, USTHB University,
Algiers, Algeria
{aalioua, smoussaoui}@usthb.dz
[2] DRIVE Labs, University of Burgundy,
Nevers, France
Sidi-Mohammed.Senouci@u-bourgogne.fr

**Abstract.** In the last few years, the emerging network architecture paradigm of Software-Defined Networking (SDN), has become one of the most important technology to manage large scale networks such as Vehicular Ad-hoc Networks (VANETs). Recently, several works have shown interest in the use of SDN paradigm in VANETs. SDN brings flexibility, scalability and management facility to current VANETs. However, almost all of proposed Software-Defined VANET (SDVN) architectures are infrastructure-based. This paper will focus on how to enable SDN in infrastructure-less vehicular environments. For this aim, we propose a novel distributed SDN-based architecture for uncovered infrastructure-less vehicular scenarios. It is a scalable cluster-based architecture with distributed mobile controllers and a reliable fallback recovery mechanism based on self-organized clustering and failure anticipation.

**Keywords**: Vehicular Ad-hoc Networks, Infrastructure-less Zones, Software-Defined Networking, Distributed Control, Mobile Controllers, Clustering

## 1 Introduction

Actually, people spend more and more time in transportation, whether in personnel vehicles or public transport. Thus, vehicles have become an important part of peoples travel experience. In this context, Intelligent Transportation System (ITS) and more specially Vehicular Ad-hoc Networks (VANETs) have attracted in the last past decade lots of interests for the purpose of improving travelling safety, comfort and efficiency via enabling communication between vehicles in an infrastructure-less vehicle-to-vehicle (V2V) mode and / or between vehicles and infrastructure in an infrastructure-based vehicle-to-infrastructure (V2I) mode.

Nowadays, VANET architectures suffer from scalability issues since it is very difficult to deploy services in a large-scale, dense and dynamic topology [1]. These architectures are rigid, difficult to manage and suffer from a lack of flexibility and adaptability in control. Hence, it is hard to choose the adequate solution to use, according to the actual context, because of the diversity of deployment environments and the large panoply of solutions that are generally adapted only in a certain context and particular situation. These constraints limit system functionality, slow down creativity and often lead to under-exploitation of network resources. Therefore, current VANET architectures cannot efficiently deal with these increasing challenges and the need of a new flexible and scalable VANET architecture become an absolute requirement.

In the last few years, the emerging network architecture paradigm of Software-Defined Networking (SDN) has become one of the most important technologies to manage large scale networks. SDN has been proposed as an attractive and promising paradigm to address the previous VANET architecture challenges. SDN is mainly based on a physical separation between control plane (network management features) and data plane (data forwarding features) and a logically centralized control and intelligence in a software controller. Other remaining equipment becomes simple data transmitter \ receiver with minimal intelligence. OpenFlow [2] is the most used standard for communication between the control plane and data plane. OpenFlow defines two types of network equipment, the controller that centralizes the network intelligence and vSwitchs that ensure only data forwarding. The controller handles the vSwitchs via installing flow entries in flow tables.

Given the growing popularity of SDN, researchers are increasingly exploring the possibility of integrating SDN in VANETs. Recently, some works have shown interest in the use of the SDN paradigm in VANETs and propose Software-Defined Vehicular Network (SDVN) architectures [1], [3–15]. These works have shown that SDN can be used to bring flexibility, scalability and programmability to VANETs, exploit the available network resources more efficiently and introduce new services in current vehicular networks. However, almost all of the proposed SDVN architectures are infrastructure-based and generally propose to host the controller somewhere on the fixed infrastructure of vehicular networks. Unfortunately, the total coverage of fixed infrastructure is not yet reached in current VANET systems. Therefore, uncovered infrastructure-less VANET zones already exist and they must be taken into consideration in the design of future SDVN architectures.

In this paper, we propose a novel distributed cluster-based architecture to enable SDN in infrastructure-less VANET environments with mobile controllers and an efficient fallback recovery mechanism based on self-organized clustering and failure anticipation. Integrating SDN in such uncovered areas is much more difficult given the dynamic sparse topology, the absence of the infrastructure support, and the vehicles' autonomous nature. To the best of our knowledge, this is the first work that integrates SDN in infrastructure-less VANET environments.

Our key contributions can be summarized as follows:
- We propose a new distributed scalable SDN-based architecture for infrastructure-less VANET,
- We present a novel kind of mobile multi-controllers, installed close to mobile vehicles to ensure a reasonable end-to-end delay and better support delay-sensitive-services,
- We furnish an efficient fallback recovery mechanism based on self-organized clustering and controller failure anticipation.

The rest of this paper is organized as follows. First, we provide an overview of existing related SDVN architectures in Section 2. In Section 3, the proposed distributed SDN-based infrastructure-less VANET architecture is presented and the fallback recovery mechanism is described. Section 4 brings two possible use-cases for the proposed architecture. Based on experimental evaluations, we demonstrate in Section 5 the reliability and the efficiency of our proposition. Finally, the conclusion is drawn in Section 6.

## 2 Related Work

Since the first work proposed in 2014 by Ku et al. [1] that explores how to integrate SDN in VANET scenarios, researchers investigated more and more how to benefit from SDN advantages to improve the performance of current VANET architectures. Recently, some SDVN works [1], [3–10] propose SDN-based architecture with the use of one controller generally hosted somewhere on the fixed infrastructure to handle the entire network. However, this assumption seems clearly impracticable in such dynamic, dense and large networks. In fact, this generates a high end-to-end delay especially when the distance between vehicles and controller is too large, which makes this solution not suitable for most delay-sensitive services. There is also a high risk of controller bottleneck and control overhead in case of a huge number of vehicle requests. Given these limitations, authors in [11–15] propose to use multiple controllers instead of a single controller and each controller handles a part of the network. The use of multiple distributed controllers can achieve scalability and reliability even in dense and heavy data loads. Much better, some of the previous works [5], [12–14] propose to install controllers (all or some) in the edge of the network, the closest as possible to vehicles to guarantee a reasonable end-to-end delay and satisfy the delay-sensitive application requirements. Furthermore, the fully centralized control of SDN presents a serious risk of reliability and security if the controller is unreachable especially if the system control is based on one unique controller, even worse with the intermittent and unstable nature of wireless connections. For this aim, works in [1], [3], [9], [12], [14] propose to use a fallback recovery mechanism as a backup solution if the controller is inaccessible.

Almost all proposed SDVN architectures are infrastructure-based and then host their controllers somewhere on the fixed infrastructure and no alternative was proposed for infrastructure-less zones, when the fixed infrastructure is totally absent or the coverage is not available. If we admit that the total and full coverage of fixed infrastructure is not yet reached in VANET system even with the integration of new emerging heterogeneous technologies as cellular technologies. For example inside tunnels when the coverage is not reachable and some rural areas where the fixed infrastructure is totally absent either because the deployment is very difficult, costly or not profitable. Thus, uncovered infrastructure-less VANET areas still exist and must be taken into consideration in the design of future SDVN architectures. We propose in the next section a distributed cluster-based SDN architecture for infrastructure-less VANET scenarios with multiple mobile controllers.

## 3 Proposed Architecture

In this section, we present our novel distributed multi-hop SDN-based architecture for infrastructure-less VANETs that uses only V2V communications via IEEE 802.11p, called *distributed Software-Defined infrastructure-less Vehicular Network (dSDiVN)*. dSDiVN uses the emerging concept of SDN to introduce flexibility, facility and scalability to the network. However, as it is well known, SDN relies on a centralized control for network management, which seems impracticable in large scale networks such as VANETs. Thus, dSDiVN proposes to combine SDN with clustering technique to partition the network and assign for each partition a dedicated controller. Hence, dSDiVN is based on a logically centralized, but physically distributed multi-hop control plane, which can benefit from the scalability and reliability of the distributed architecture while preserving the simplicity of the centralized management. dSDiVN uses multiple mobile controllers that interact each other and work together to get a global view of the network state. Partitioning the network makes it more stable, smaller and less dynamic for a vehicle, and can reduce overhead and latency.

The dSDiVN architecture is detailed below.

### 3.1 dSDiVN System Architecture

dSDiVN is based on a combination of two emerging network paradigms: SDN and clustering. Indeed, the control logic in clustering technique when the cluster head centralizes the cluster intelligence and the cluster members have a minimal intelligence, is very similar to that used by SDN paradigm. In this context, dSDiVN involves to: *(i)* organizing and partitioning the network according to certain criteria in partitions (*i.e.,* segments) using the clustering technique, *(ii)* grouping vehicles that are situated in the same geographic area (*i.e.,* segment) and which have similar characteristics (position, velocity, direction, etc.) at the same virtual group (i.e., cluster), *(iii)* choosing a leader (*i.e.,* cluster head) for each partition, *(iv)* deploying a local controller on each partition leader (*i.e.,* cluster head), and *(v)* connecting the adjacent controllers to build a backhaul for network control and enforce global policies. By partitioning the network and using distributed controllers, dSDiVN can better deal with scalability, handles easily increasing load, introduces efficiently specific new services to a particular cluster and effectively offers more reliability.

For cluster's management and maintenance, dSDiVN adopts and adapts the clustering algorithm in [16] based on CGP [17], which is a distributed multi-hop geographic clustering algorithm. As shown in Fig.1, the uncovered road is divided into equal size segments of 150 m each (half of IEEE 802.11p coverage area) to ensure that adjacent controllers on cluster heads always be reachable to each other. Each segment represents a virtual cluster (called *Software-Defined domain, SD-domain*), that regroups all mobile vehicles (called *Software-Defined mobile vehicle, SD-vehicle*) which roll in the same direction and are situated in the same SD-domain. The elected cluster head (called *SD-domain head, SD-DH*) will be the SD-vehicle that has the longest time to life in the SD-domain (*i.e.,* the longest SD-domain travel time) to minimize the cluster maintenance overhead. Upon elected, each SD-vehicle domain head enables its local mobile controller and starts managing a backup candidate list in order to anticipate its potential failure and prepare the recovery controller. The identifier of the best candidate (*i.e.,* recovery controller) is sent periodically to all cluster members (called *SD-domain members, SD-DM*) via flow rules and stored in the flow table. For that, we propose to add a novel field to the flow table entry for storing the identifier. A replication of local mobile controller knowledge base is compressed and backed-up periodically on the recovery controller to allow fast service resume if the controller failed. Adjacent local mobile controllers are connected with each other via IEEE 802.11p to build a control backhaul for dSDiVN, see Fig. 1.

Each local mobile controller maintains a local view of the network state of its SD-domain. A global view of the network state can be obtained by exchanging the local view of adjacent controllers. The local mobile controller is considered in control as a master of its SD-domain members where it collects data from the multiple members (slaves), and equivalent to its mobile neighbor controllers where all the controllers collaborate to get a global view of the entire network.

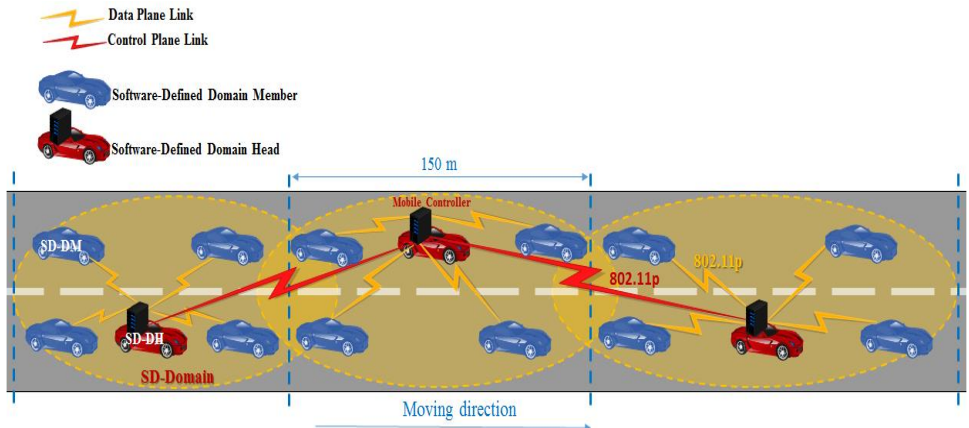

**Fig. 1.** dSDiVN System architecture.

### 3.2 dSDiVN SDN Architecture

dSDiVN is based on the three layers of SDN architecture, as illustrated in Fig. 2:
1) **Data plane layer**: It consists of all SD-vehicle SD-domain members that only perform collection and forwarding of data information,
2) **Control plane layer**: It consists of all local mobile controllers deployed on the SD-vehicle SD-domain heads that centralize the network control,
3) **Service and application layer**: It contains all the services and applications such as routing, security and QoS services. To minimize the cost of installing new services, the service is initially installed in one controller and after that communicated hop by hop to the adjacent controllers,
4) **Communication interfaces**: OpenFlow is an IP-oriented protocol and there is not a no-IP version compatible with safety VANET application features. Also, there are no standardized SDN interfaces for directly integrating SDN into VANETs. Therefore, we propose to use: *(i)* a customized version of OpenFlow protocol adapted to V2V communications as southbound API to communicate between the control plane and the data plane and, *(ii)* a customized interface as northbound API to communicate between the control plane and applications [9].

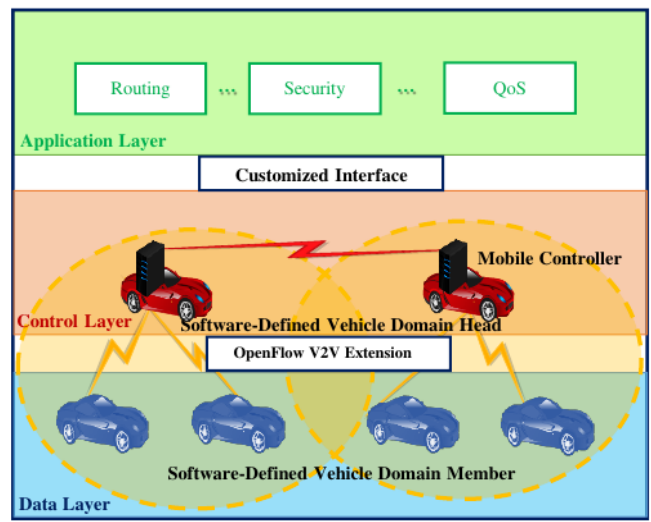

**Fig. 2.** dSDiVN 3-tier SDN architecture.

In dSDiVN, all SDN components are implemented on the unique hardware component, *i.e.,* the wireless mobile vehicle that the internal architecture is detailed below.

### 3.3 Software-Defined Wireless Mobile Vehicle

A software-defined wireless mobile vehicle (SD-vehicle) is a traditional vehicle with an additional SDN module which consists of hardware and software resources that allow SDN to function on the mobile vehicle. From the hardware side, it consists of a computing and storage unit that offers the SDN platform to install different services (to limit hardware modification, it is preferable to reuse, if possible, the available resources of the OBUs - on board units). From the software side, it consists of basic SDN components: SDN operating system, virtual machines on the hypervisor, network services, etc. Moreover, dSDiVN defines two main software SDN components, as illustrated in Fig. 3:

1) *The local mobile controller (simplified mobile controller)*: It is initially in standby mode and it is enabled when the hosting SD-vehicle is selected as an SD-domain head. It is known as mobile because it is implemented on a mobile vehicle and can migrate from an SD-domain head to another, when it hosting SD-vehicle leave an SD-domain. The mobile controller centralizes the intelligence and controls all the vehicles of its SD-domain. To the best of our knowledge, we are the first that propose a mobile controller in an SDVN architecture.

2) *The monitoring and collection agent*: it ensures data forwarding and monitoring of SD-vehicle parameters (*e.g.,* position, velocity, direction, etc.). This monitored information is periodically communicated to the mobile controller. The local monitoring and collect agent receives and executes control directives from its mobile controller via flow rule entries installed in its flow table.

The architecture of the SD-vehicle is implemented at the *facilities layer* of the standard *ISO CALM (Communication Architecture for Land Mobile)*, which assumes the existence of multiple wireless interfaces in a vehicle. In dSDiVN, each SD-vehicle has several communication interfaces: *(i)* two broadband wireless interfaces DSRC (*i.e.,* IEEE 802.11p), to avoid interferences and separate control traffic from data traffic: one for the V2V control plane communication between SD-domain members and the mobile controller and the other for the data plane V2V communication between vehicle SD-domain members and, *(ii)* a wideband wireless cellular interface (LTE/4G), initially disabled and enabled when the coverage of fixed infrastructure is available, as illustrated in Fig. 3.

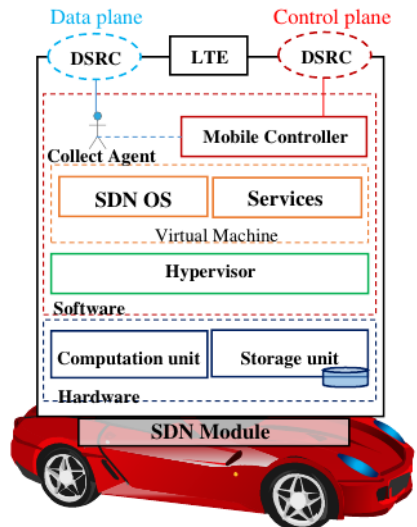

**Fig. 3.** Internal architecture of software-defined mobile vehicle.

The fallback recovery mechanism of dSDiVN is presented below.

### 3.4 Fallback Recovery Mechanism

In a fully centralized architecture such as the one that adopts SDN, all system reliability relies on the central controller. Therefore, it is necessary to envisage a back recovery mechanism if this central controller is unreachable [1]. In [1], [14], authors propose to switch to traditional VANET routing when the connection with the controller is lost. This may increase the complexity of software and hardware design [12]. In [9], authors use trajectory prediction to pre-install entries in the flow table that can be used for a short period of time when the controller fails but this initiative does not envisage any solution if the controller failure lasts after the end of the entry. In [12], two types of hierarchical controllers are used, and if the high-level controller fails, the low-level controllers take over to ensure service continuity but no solution is envisaged if one of the low-level controllers failed.

Our architecture is based on a cluster-based distributed multi-hop control system. As fallback recovery mechanism, we simply propose to profit from the self-organized clustering technique, permanently anticipating the possible failure of each controller and prepare in advance the recovery scenario. Thus, if a mobile controller fails, the pre-prepared recovery controller (see section 3.A) will take over to ensure the service continuity. When the SD-vehicles are aware of the controller failure, they will start to send their requests directly to the recovery controller using the beforehand pre-installed identifier as prevention recovery solution, see Section 3.A. The recovery controller when enabled, use the knowledge base replication to immediately start to respond the SD-vehicle requests.

Some possible use-cases of dSDiVN are presented in the next section.

## 4. dSDiVN Use Cases

In this section we present two possible use-cases of our dSDiVN architecture:
- *SDN-assisted efficient data collection and dissemination*: the data collected at each SD-vehicle level are not useful individually. However, the aggregation of all data of a geographical area allows having a vision on the network state. By separating the data plane from the control plane, mobile controllers can centralize the collection of data from various sensors installed on SD-vehicles, each in its SD-domain. Unlike traditional networks, mobile controllers can handle the collected data in a more informed way to extract useful information and improve the system decision. Also, by collaborating with each other, mobile controllers can build a global view of current network state and can choose the most optimal path to disseminate/route this extracted information.
- *SDN-assisted VANET safety applications*: dSDiVN uses only V2V based IEEE 802.11p communications, which are more suitable for safety applications. Moreover, mobile controllers based on the information observed of current traffic conditions can collaborate each other to: *(i)* ensure better persistence and availability of emergency alert messages by accurately defining the danger zone extent (size) and the duration of the emergency alert, *(ii)* choose the fastest path to disseminate the alert message, and *(iii)* treat the emergency traffic with more priority in reservation of specific frequencies and channels over the remaining normal traffic.

## 5. Numerical Results

In this section, we describe our simulation configuration, metrics and results. The simulation model is built based on the system architecture described in Section 3.A, and it is implemented using the network simulator NS3 [18] and the traffic simulator SUMO [19]. The aim of the simulations is to evaluate the reliability of the fallback recovery mechanism and the effect of controller distance on flow rule installation time.

For simulation, we consider an infrastructure-less VANET scenario where the network is deployed in 1000 x 1000 m area (a cell of Manhattan Grid). The road is divided into equal segments of 150 m each, see Fig. 1. Node density is 200 nodes. Each vehicle has IEEE 802.11p based interface with a transmission range up to 300 m and a velocity between 10 and 30 m/s. The simulation time is 120 seconds and the packet generation rate is 5 packets/s. The performance metrics we used are:

- *Flow rule installation time*: it represents the elapsed time since an SD-vehicle requests the controller for a new flow rule and the time when the flow rule is installed in the SD-vehicle flow table,
- *Packet delivery ratio*: is the ratio of packets successfully received to the total sent.

### 5.1 Controller Failure

In this evaluation, we study how dSDiVN with the proposed fallback recovery mechanism (multi-controllers cluster-based with failure anticipation and recovery controller pre-preparation) reacts to the controller failure. We focus on one cluster and we simulate a controller failure of 5 seconds at the $61^{st}$ second. Afterwards, we compare the dSDiVN fallback recovery mechanism (we simplified dSDiVN) with that of self-organized cluster-based of dSDiVN without fallback recovery (we simplified self-organized) and with works [3–8], [10], that use one controller without fallback recovery mechanism (we simplified no-back-recovery) according to the packet delivery ratio. The comparison results are illustrated in Fig. 4.

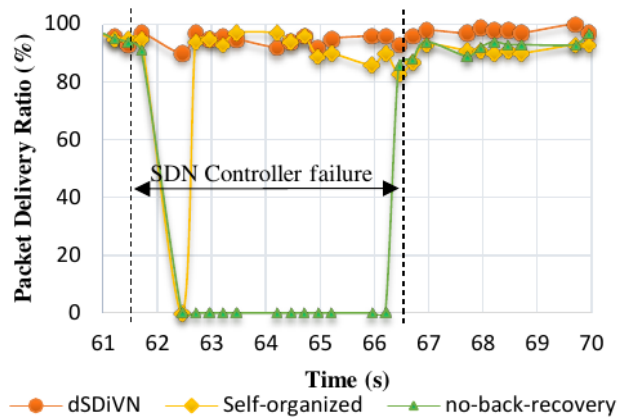

**Fig. 4.** Controller failure.

As illustrated in Fig. 4, we can see that just when the controller fails, the packet delivery ratio in no-fall-recovery starts to decrease dramatically and it still drops until the controller resumes the service. For the case of self-organized, the delivery radio drops for a short time, which represents the time until the reelection of a new SD-domain head and the synchronization with the newly elected controller; after that, the system resumes a good delivery ratio. This behavior can be justified by the fact that in the SDVN solution, the reliability of the system resides on one central controller and as soon as the controller fails, it stops installing flow rules in flow tables; therefore, the service is interrupted. Much more resistant, the packet delivery ratio in dSDiVN undergoes a very slight effect, thanks to the efficient preventive fallback recovery mechanism based on distributed multi-controllers, self-organized clustering and failure anticipation. When the controller fails, SD-vehicles can directly synchronize and direct their requests to the pre-selected recovery controller using the pre-installed identifier, see Section 3.A.

This evaluation demonstrates the reliability and the efficiency of our fallback recovery mechanism and confirms the result in [1]: the use of a fallback recovery mechanism is primordial when operating the centralized SDN-based control in VANET especially when the system control lies on one central controller, more worst with the intermittent wireless link nature.

### 5.2 Effect of Controller Distance on Flow Rule Installation Time

In this evaluation, we aim to study the effect of controller distance, which represents the physical distance between the controller and SD-vehicles, on the flow rule installation time. Thus, we use a simple scenario when an SD-vehicle sends a request to its controller and we have varied the distance between the SD-vehicle and the controller for different request packet sizes.

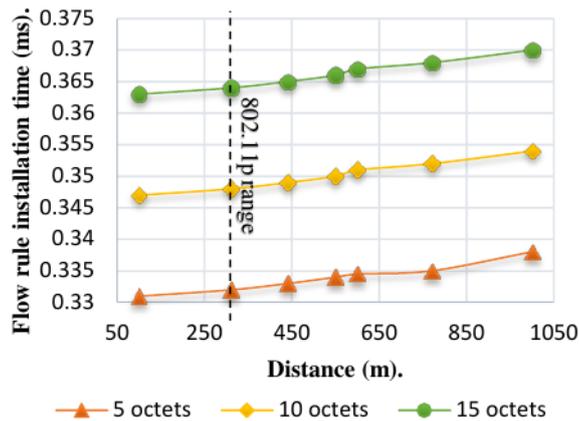

**Fig. 5.** Controller distance vs. flow rule installation time.

Fig. 5 shows that flow rule installation time increases with the increase of the distance between the controller and the SD-vehicle and even with the increase of packet size. This evaluation is performed in a cell of Manhattan Grid with 1000 m using LTE communication. However, the result can be generalized for large scale scenarios in infrastructure-based areas. In this case, the result can be too worst and the system cannot ensure a packet delivery ratio that satisfies the requirements of delay-sensitive services especially when the distance reaches hundreds of kilometers.

From this evaluation, we can conclude that the controller should be installed the closet as possible to vehicles to guarantee the suitable flow rule installation time of delay-sensitive services and react rapidly to real-time events.

In our architecture, the controllers are installed on the SD-domain head vehicles (the farthest distance of a controller is 150 m) and communicate with forwarding vehicles via V2V and IEEE 802.11p (more adequate to satisfy safety service requirements), which allows dSDiVN to ensure a good and adequate flow rule installation time and a better support of delay-sensitive services and real-time events.

## 6. Conclusion

In this paper, we propose dSDiVN, distributed Software-Defined infrastructure-less Vehicular Network, a novel distributed multi-hop SDN-based clustered architecture for infrastructure-less vehicular networks with mobile multi-controllers and a reliable fallback recovery mechanism. Our dSDiVN bridges the gap of no SDN-based architecture in infrastructure-less VANET zones. Numerical results demonstrate that the reliability of our fallback recovery mechanism and the negative effect that represents the far distance of controller on the flow rule installation time which is the main requirement of the VANET safety applications.

The sparse network topology that results in partitioning of the network rests an open challenge in front of integrating SDN in infrastructure-less VANET areas. As short-term perspective of this work, we plan to extend our architecture to deal with the network partitioning problem by introducing drones to connect isolated mobile controllers, as we plan to do more advanced experimentations to proof the feasibility and the efficiency of the proposed architecture.